\documentclass[twocolumn,showpacs,superscriptaddress,preprintnumbers,amsmath,amssymb,prl,aps]{revtex4-1}
\usepackage{graphicx}
\usepackage{dcolumn}
\usepackage{bm}

\begin{document}

\title{Spin-flip Raman scattering of the $\Gamma$-X mixed exciton in indirect band-gap (In,Al)As/AlAs quantum dots}

\author{J.~Debus}
\email[Corresponding author: ]{joerg.debus@tu-dortmund.de}
\affiliation{Experimentelle Physik 2, Technische Universit\"at Dortmund, 44227 Dortmund, Germany}

\author{T.~S.~Shamirzaev}
\affiliation{Institute of Semiconductor Physics, Russian Academy of Sciences, 630090 Novosibirsk, Russia} \affiliation{Ural Federal University, 620002 Yekaterinburg, Russia}

\author{D.~Dunker}
\affiliation{Experimentelle Physik 2, Technische Universit\"at Dortmund, 44227 Dortmund, Germany}

\author{V.~F.~Sapega}
\affiliation{Ioffe Physical-Technical Institute, Russian Academy of Sciences, 194021 St. Petersburg, Russia} \affiliation{Spin Optics Laboratory, St. Petersburg State University, 198504 St. Petersburg, Russia}

\author{E.~L.~Ivchenko}
\affiliation{Ioffe Physical-Technical Institute, Russian Academy of Sciences, 194021 St. Petersburg, Russia}

\author{D.~R.~Yakovlev}
\affiliation{Experimentelle Physik 2, Technische Universit\"at Dortmund, 44227 Dortmund, Germany} \affiliation{Ioffe Physical-Technical Institute, Russian Academy of Sciences, 194021 St. Petersburg, Russia}

\author{A.~I.~Toropov}
\affiliation{Institute of Semiconductor Physics, Russian Academy of Sciences, 630090 Novosibirsk, Russia}

\author{M.~Bayer}
\affiliation{Experimentelle Physik 2, Technische Universit\"at Dortmund, 44227 Dortmund, Germany}

\begin{abstract}
The band structure of type-I (In,Al)As/AlAs quantum dots with band gap energy exceeding $1.63$~eV is indirect in momentum space, leading to long-lived exciton states with potential applications in quantum information. Optical access to these excitons is provided by mixing of the $\Gamma$- and X-conduction band valleys, from which control of their spin states can be gained. This access is used here for studying the exciton spin-level structure by resonant spin-flip Raman scattering, allowing us to accurately measure the anisotropic hole and isotropic electron $g$ factors. The spin-flip mechanisms for the indirect exciton and its constituents as well as the underlying optical selection rules are determined. The spin-flip intensity is a reliable measure of the strength of $\Gamma$-X-valley mixing, as evidenced by both experiment and theory.
\end{abstract}


\maketitle

While semiconductor quantum dots (QDs) have been established as efficient light emitters and detectors in optoelectronics~\cite{QDbook}, other applications are only prospective so far. Particular examples are implementations in spin electronics and quantum information technologies. For these purposes, the QDs are typically loaded with resident carriers whose spins are well protected from relaxation by the three-dimensional confinement~\cite{Cortez,Kammerer}. In this context, exciton complexes are often used for spin manipulation~\cite{Kroutvar,Senes}, but are considered less promising as information carriers. This reservation is primarily related to the limited exciton lifetime of about a nanosecond~\cite{Yu}, which is too short to provide sufficient coherent manipulation~\cite{Awschalom}. This situation may change if the exciton lifetime could be extended significantly.

Interesting, but technologically challenging in this respect is the placement of QDs in photonic crystals, in which their radiative decay could be suppressed~\cite{Vuckovic,Kress}. Alternatively, instead of the bright excitons dark excitons with lifetimes in the $\mu$s-range may be used as information carriers, which however complicates the direct optical manipulation, so that manipulation through the biexciton state can be realized~\cite{Gershoni}. Another possibility is the realization of QDs with a band gap that is indirect in real or momentum space. Here, we focus on self-assembled (In,Al)As/AlAs QDs, in which a crossover of the lowest conduction band states between the $\Gamma$- and X-valley occurs~\cite{Shamirzaevb}, depending on the dot size. This crossover is reflected by the lifetime of the corresponding exciton, which is formed by a $\Gamma$-valley heavy-hole and a $\Gamma$- and/or X-valley electron. Both carriers are spatially located within the QD (type-I band alignment). If the $\Gamma$- and X-electron states become admixed, the lifetime of that exciton can be as long as hundreds of $\mu$s~\cite{Shamirzaevnew}, which may allow sufficient manipulation within this time span.

Studying and manipulating an X-valley electron and, in turn, an exciton that is indirect in momentum space by optical techniques poses, however, a significant problem: the associated optical transitions are forbidden in bulk crystals and are only weakly allowed in QDs due to breaking of the translational symmetry. This limitation may be bypassed by utilizing state mixing of the direct and indirect conduction-band minima in the (In,Al)As/AlAs QDs. One appealing optical technique, resonant spin-flip Raman scattering (SFRS), may then allow one to study the spin properties of the indirect exciton. SFRS spectroscopy, however, is not only a powerful tool to probe spins by measuring $g$ factors, but is also able to exploit spin interactions to orient spins. It has been successfully applied to quantum wells~\cite{Debus,Koudinov} and nanocrystals~\cite{Sirenko}, and has been suggested for direct-gap QD studies~\cite{Puls}.

In this Letter we demonstrate that indirect-in-momentum-space excitons can be addressed optically by SFRS in an ensemble of undoped (In,Al)As/AlAs QDs. We use SFRS to characterize the $\Gamma$-X-valley electron state mixing. It provides access to the fine structure of the indirect exciton and allows us to measure the $g$ factor tensor components of the indirect exciton and its constituents. The mechanisms of their spin-flips and the optical selection rules are determined. The electron spin-flip energy and probability are theoretically modeled by considering an acoustic phonon scattering process including the exciton lifetimes and $\Gamma$-X-mixing parameters. 

The studied structure contains 20 layers of undoped (In,Al)As/AlAs QDs grown by molecular-beam epitaxy on a (001)-oriented GaAs substrate. The density of the lens-shaped QDs with an average diameter of $15$~nm and height of $4$~nm is about $3 \times 10^{10}$~cm$^{-2}$ in each layer. The QD layers are separated from each other by $20$-nm-thick AlAs barriers, which prevent an electronic coupling between QDs in adjacent layers. SFRS spectra are measured in the backscattering geometry at a temperature of $1.8$~K with circular or linear polarization for the incident and scattered light~\cite{Debus}. The angle $\theta$ between the magnetic field $\textbf{B}$ and the QD growth axis $\mathbf{z}$ is varied between 0$^\circ$ (Faraday geometry) and 90$^\circ$ (Voigt geometry).

\begin{figure}[t]
\centering
\includegraphics[width=8.5cm]{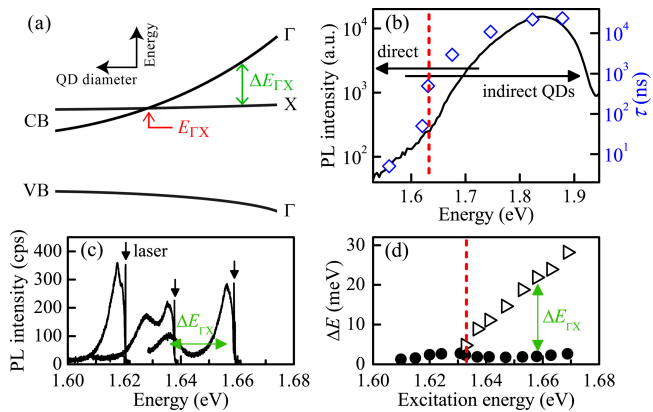}
\caption{\label{fig1} (Color online) (a) Band alignment in (In,Al)As/AlAs QDs as function of dot diameter for the valence (VB) and conduction (CB) bands. The energy difference between $\Gamma$- and X-electron states is denoted by $\Delta E_{\Gamma \text{X}}$. For a particular diameter $\Gamma$-X crossover occurs, corresponding to a gap energy $E_{\Gamma \text{X}}$. (b) PL spectrum of an (In,Al)As/AlAs QD ensemble at $T=1.8$~K; excitation photon energy $E_\text{exc} = 2.33$~eV. The exciton recombination times $\tau$ across the ensemble are shown by open diamonds (right scale). (c) Resonantly excited PL of direct and indirect excitons around the $\Gamma$-X crossover. The laser photon energies are marked by arrows. (d) Difference between laser energy and the peak position of the direct (circles) and indirect (triangles) exciton PL.}
\end{figure}

Dispersion in dot size, shape, and composition within the ensemble leads to formation of (In,Al)As/AlAs QDs with different band alignments, as shown in Fig.~\ref{fig1}(a). The electron (e) ground state changes from the $\Gamma$- to the X-valley with decreasing dot diameter, while the heavy-hole (hh) ground state remains at the $\Gamma$-point. This corresponds to a change from a direct to an indirect band gap in momentum space while type-I band alignment is preserved~\cite{Shamirzaev,Shamirzaevb}: the lowest electron level arises from the X-valley in small-diameter QDs with strong quantum confinement along the growth direction. With increasing dot diameter the $\Gamma$-valley level shifts to lower energies more rapidly than the X-level, due to the smaller effective mass of $\Gamma$-valley electrons~\cite{Luo}. For a particular dot diameter the $\Gamma$- and X-electron levels intersect. The corresponding crossing energy is marked in Fig.~\ref{fig1}(a) by $E_{\Gamma \text{X}}$. Note that the quantum confinement splits the degenerate X-electron states into X$_{xy}$ and X$_z$ states with the valley main axis being perpendicular and parallel to the $z$-axis, respectively. The X$_{xy}$ state has lower energy~\cite{Shamirzaev}, we refer to it as the X-valley electron state in the following.

The coexistence of QDs with direct and indirect band gaps within the ensemble is evidenced by the spectral dependence of the radiative exciton recombination times $\tau$. As shown in Fig.~\ref{fig1}(b), the indirect QDs are characterized by long decay times in the $\mu$s-range due to the small exciton oscillator strength~\cite{Shamirzaevnew}. These indirect excitons also exhibit long longitudinal spin relaxation times of up to $200$~$\mu$s in magnetic field, see SOM. On the contrary, in the direct band-gap dots the excitons recombine within a few nanoseconds. In the $\Gamma$-X-crossover range, the exciton photoluminescence (PL) decay contains contributions from both direct and indirect excitons~\cite{commentrange}.

Further insight into the $\Gamma$-X crossing can be obtained from PL under resonant excitation, which selects only a fraction of dots in the ensemble, causing line narrowing due to reduced inhomogeneous broadening. One can see in Fig.~\ref{fig1}(c) that for low-energy excitation with $E_\text{exc} < E_{\Gamma \text{X}}$ only the largest dots hosting direct excitons are excited, thus resulting in a spectrally narrow PL line. For excitation energies exceeding $E_{\Gamma \text{X}}$, an additional broad PL line appears, which originates from indirect exciton emission. The separation $\Delta E$ of the emission line maximum from the varying laser photon energy is plotted in Fig.~\ref{fig1}(d). The direct excitons (circles) closely follow $E_\text{exc}$ with a small shift $\Delta E = (2.2\pm 0.1)$~meV. This shift arises from excitation through an acoustic phonon, which is most efficient for the phonon wavelength matching the dot size. On the other hand, the shift of the indirect exciton PL line (triangles) increases markedly and linearly with $E_\text{exc}$, as the recombination energy remains almost fixed. The meeting point of both shifts at 1.633~eV occurs at $E_{\Gamma \text{X}}$, indicated by the dashed line.

\begin{figure}[t]
\centering
\includegraphics[width=8.5cm]{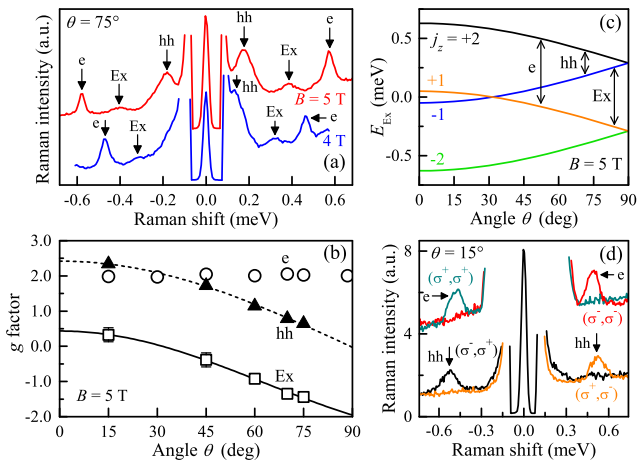}
\caption{\label{fig2} (Color online) (a) Stokes and anti-Stokes SFRS spectra for magnetic fields of 4 and 5~T in tilted geometry ($\theta = 75^\circ$) with crossed linear polarization; $T=1.8$~K. Resonant excitation close to the $\Gamma$-X crossing at 1.636~eV, laser power density at the sample is about $15$~W/cm$^2$. (b) $g$ factor angle dependence for the heavy-hole, X-valley electron and indirect exciton. Dashed line is the fit for $g_{\text{hh}}(\theta)$, solid line is the calculation for $g_{\text{Ex}}(\theta)$, see text. Note, the points $g_{\text{Ex}}(15^\circ)$ and $g_{\text{Ex}}(45^\circ)$ are evaluated from the measured e and hh $g$ factors. (c) Calculated energies of angle-dependent bright and dark exciton states at $B = 5$~T, spin-flip transitions are shown by arrows. The center-of-gravity is taken as zero. (d) Cross- and co-circularly polarized SFRS spectra measured at $B=4$~T and $E_\text{exc} = 1.644$~eV.}
\end{figure}

Now, let us study the exciton spin-level structure and spin-flip mechanisms by use of the resonant SFRS. Raman spectra recorded at magnetic fields of 4 and 5~T in a tilted geometry ($\theta = 75^\circ$), essential for the symmetry breaking required for spin flips, are shown in Fig.~\ref{fig2}(a) for excitation at the $\Gamma$-X-crossover energy. Three SFRS lines corresponding to the heavy-hole, X-valley electron and indirect exciton (Ex) are observed in the Stokes and anti-Stokes regions. Their spin-flip Raman shifts $\Delta E_{\text{SF}}$ correspond to transitions between Zeeman sublevels split by $|g| \mu_{\text{B}} B$ with the Bohr magneton $\mu_{\text{B}}$. One obtains $g$ factors of $|g_{\text{e}}^{\theta}| = 2.00 \pm 0.01$, $|g_{\text{Ex}}^{\theta}|=1.24 \pm 0.02$, and $|g_{\text{hh}}^{\theta}|=0.75 \pm 0.01$ for $\theta = 75^\circ$, see SOM for details.

The angular dependence of the $g$ factors at $B = 5$~T is demonstrated in Fig.~\ref{fig2}(b). The shift of the e-SFRS line is isotropic, $g_{\text{e}} \equiv g_{\text{e}}^\parallel = g_{\text{e}}^{\perp}$, with $|g_{\text{e}}| = 2.00 \pm 0.01$. The $g$ factor isotropy and magnitude are characteristic for X-valley electrons in indirect band-gap structures~\cite{Young,Vdovin}. Due to the large band gap at the X-point ($\approx 4.8$~eV between conduction and valence band) the spin-orbit contribution to the electron $g$ factor is vanishingly small~\cite{Yugovab}. As a result, the measured value coincides with the free electron Land\'e factor.

The angular dependent $g$ factors in Fig.~\ref{fig2}(b) are assigned to the heavy-hole and indirect exciton. The hh $g$ factor for a particular field direction is determined by its tensor components along and normal to the growth direction through $g_{\text{hh}}(\theta)=[(g_{\text{hh}}^{\parallel} \cos \theta)^{2} + (g_{\text{hh}}^{\perp} \sin \theta)^{2} ]^{1/2}$. As seen from the corresponding fit (dashed line), $g_{\text{hh}}(\theta)$ describes well the experimental data with $g_{\text{hh}}^{\parallel} = 2.42 \pm 0.05$ and $g_{\text{hh}}^{\perp} = 0.03 \pm 0.05$. The small transverse $g$ factor indicates weak mixing of light-hole (lh) and heavy-hole at the $\Gamma$-point compared to, e.g., (In,Ga)As QDs.

\begin{figure}[t]
\centering
\includegraphics[width=6.7cm]{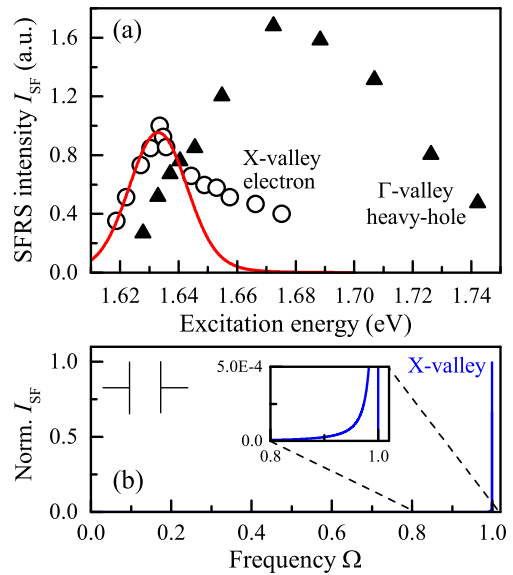}
\caption{\label{fig3} (Color online) (a) e- and hh-SFRS resonance profiles measured at $B = 5$~T, $T = 1.8$~K, $\theta = 75^\circ$. (b) Simulated Raman spectrum for an X-valley state, after Eq.~(\ref{ramanspectheo}), and scheme of spectral bandpass of monochromator slits. Inset indicates the smooth tail of the X-valley line for $\Omega < 1$.}
\end{figure}

Bearing in mind the isotropy of $g_{\text{e}}$ and the positive sign of $g_{\text{hh}}^{\parallel}$, we can evaluate the indirect exciton $g$ factor from $g_{\text{Ex}}(\theta) = g_{\text{hh}}^{\parallel} \cos \theta - g_{\text{e}}$. The calculated dependence for $g_{\text{Ex}}(\theta)$ shown by the solid line in Fig.~\ref{fig2}(b) is in good accord with the data. The following exciton $g$ factor values are obtained: $g_{\text{Ex}}^{\parallel} = 0.43 \pm 0.08$ and $g_{\text{Ex}}^{\perp} = -1.95 \pm 0.08$. The positive sign of $g_{\text{Ex}}^{\parallel}$ is supported by the magnetic-field-induced circular polarization of the QD photoluminescence, see SOM. The resulting exciton fine-structure pattern as function of the tilt angle is plotted in Fig.~\ref{fig2}(c) together with the spin-flip transitions. The values are calculated using the magnetic-field dependent Hamiltonian for a $D_{2d}$ QD point symmetry. Details of the calculations are given in the SOM. The transitions between the bright exciton states with $|j_z| = 1$ require simultaneous reversals of the electron and hole spins, either due to one-phonon or two-phonon processes~\cite{Sapega}. In the one-phonon process the simultaneous flip occurs via the heavy- and light-hole-exciton mixing owing to the interplay of exchange interactions and lattice deformations. The two-phonon exciton spin-flip is a double-quantum transition with a virtual intermediate state, which does not require exchange interaction. Hence, the spin of the indirect exciton is very likely flipped by the two phonon-process, while the one-phonon process is less probable due to the weak hh-lh mixing.

As depicted in Fig.~\ref{fig2}(d) for a close-to-Faraday geometry with $\theta = 15^\circ$~\cite{Faradaygeom}, the hh-SFRS is observed for crossed circular polarizations in Stokes and anti-Stokes regions, while the e-SFRS line, having a smaller Raman shift, is present in co-polarized configurations. The hh and e spins are scattered by acoustic phonons, where dark indirect excitons are the intermediate scattering states. The $\Gamma$-X-valley mixing and particularly the mixing of the electron spin states in tilted geometry allow both SFRS processes. Detailed scattering schemes can be found in the SOM.

The intensities of the e- and hh-SFRS lines that characterize the efficiencies of the Raman scattering processes are plotted in Fig.~\ref{fig3}(a) as the laser photon energy $E_{\rm exc}$ is tuned across the QD ensemble. The spectral profile of the hh-SFRS intensity is much broader than that for electrons, and has a maximum at about $1.685$~eV. However, its width is narrower than that of the ensemble PL spectrum. The profile can be explained by Raman scattering involving the direct exciton state. The spectral density of QDs shapes the low-energy side of the profile. The decrease at the high-energy side is due to shortening of the direct exciton lifetime caused by electron scattering from the $\Gamma$- to the X-valley. This process becomes efficient when the energy of the $\Gamma$-valley exceeds that of the X-valley by the longitudinal-optical phonon energy, which is 30~meV for InAs and 49~meV for AlAs phonons~\cite{HPL}.

The X-valley electron SFRS intensity has a sharp maximum at $E_{\rm max}$ = 1.633~eV, which is the crossing energy $E_{\Gamma \text{X}}$ of the $\Gamma$- and X-electron valleys. Qualitatively, this can be understood by taking into account that one-photon excitation of a pure indirect exciton is forbidden, but can be achieved by mixing the direct exciton with the indirect one. This admixture is provided by mixing of the $\Gamma$- and X-electrons. The X-valley electron SFRS intensity is expected to be maximum when the $\Gamma$- and X-valleys are in resonance.

For in-depth understanding of the SFRS process involving the $\Gamma$-X mixed exciton states we propose a model to calculate the e-SFRS spectrum including the resonant intensity profile. We consider the QDs as an ensemble of two-level systems with wave function $\Psi = C_{\Gamma} |\Gamma \rangle + C_{\text{X}} | \text{X} \rangle$, where the coefficients $C_{\Gamma (\text{X})}$ are determined by the energy difference $\Delta \equiv \Delta E_{\Gamma \text{X}} = E_{\Gamma} - E_{\text{X}}$ between the lowest $\Gamma$- and X-levels as well as the matrix element $V_{\Gamma \text{X}}$ of the $\Gamma$-X-level coupling, see SOM for details. The $g$ factor is given by $g_{\text{e}} \equiv g(\Delta) = g_{_\Gamma} |C_{\Gamma}|^2 + g_{_\text{X}} |C_{\text{X}}|^2$, where $g_{_\Gamma}$ and $g_{_\text{X}}$ are the single-valley $g$ factors~\cite{comment1}. In the following, among the two split states we only consider the one with $|C_{\text{X}}| > |C_{\Gamma}|$ because the other state does not contribute notably to the e-SFRS line in Fig.~\ref{fig2}(a). In order to express the ensemble character with its different dot sizes and shapes, the $\Gamma$-X-level splitting shall be now taken as the sum of the average value $\bar{\Delta}(E_{\text{exc}})$ plus a random value $\tilde{\Delta}$ with Gaussian distribution $F(\tilde{\Delta})$: $D = |\bar\Delta(E_{\text{exc}}) + \tilde \Delta|$.

Eventually, one obtains for the e-SFRS intensity
\begin{multline} \label{resotheory}
I_{\text{SF}}(E_{\text{exc}}) \propto \int\limits_{-\infty}^\infty d \tilde \Delta F(\tilde \Delta) \left[ \frac{\sqrt{D^2 + \delta^2} - D}{ \left( 1+2\alpha \right ) \sqrt{D^2 + \delta^2} - D} \right ]^2  \times \\
\times \left( 1+ \frac{D}{\sqrt{D^2+\delta^2}} \right)^N
\end{multline}
with $\delta = 2V_{\Gamma \text{X}}$ and $\alpha = \tau_{_\Gamma}/\tau_{\text{nr}}$, where $\tau_{\text{nr}}$ is the non-radiative exciton lifetime and $\tau_{_\Gamma}$ is the radiative lifetime of the $\Gamma$-valley exciton. Following Ref.~\onlinecite{Nazarov}, $N \approx 3 - 5$, which accounts for the one-phonon assisted electron spin-flip process in QDs. For the average splitting we assume, in accordance with Fig.~\ref{fig1}(d), a linear dependence $\bar{\Delta} = \eta (E_{\text{exc}}-E_{\text{max}})$, where $E_{\text{max}} \equiv E_{\Gamma \text{X}}$ is the incident photon energy at which the photoexcited $\Gamma$- and X-levels merge on average, and take $\eta = 0.65$ from the slope of the dependence given by the triangles in this figure.

Equation~(\ref{resotheory}) describes well the experimental e-SFRS intensity data, as shown by the red curve for $N=5$ in Fig.~\ref{fig3}(a). From the simulation we obtain accurate values for the involved parameters, in particular $\delta=0.8$~meV, and also $\Delta_0=10$~meV as well as $\alpha \leq 10^{-2}$~\cite{remark}. The high-energy tail observed in the resonance profile is caused by a more complex distribution of QD sizes than the assumed Gaussian. Nevertheless, Eq.~(\ref{resotheory}) provides a reliable way to estimate the strength of the $\Gamma$-X coupling by $V_{\Gamma \text{X}} = \delta/2 = 0.4$~meV for the studied QD ensemble.

In the range of strong $\Gamma$-X mixing one can calculate the SFRS spectrum by integration of Eq.~(\ref{resotheory}) with the delta-function $\delta(\Omega - |C_{\text{X}} (\tilde{\Delta})|^2)$ in the integrand, where $\Omega = (E_{\rm exc} - \hbar \omega_{\text{f}})/g_{_\text{X}} \mu_{\text{B}} B$ is the difference between the incident and final photon energies normalized to the electron Zeeman splitting  $g_{_\text{X}} \mu_{\text{B}} B$ of the pure X-state. The integration results in the Raman spectrum intensity
\begin{eqnarray} \label{ramanspectheo}
&&I_{\text{SF}}(\Omega) \propto \frac{{\rm e}^{-\Delta_+^2/\Delta_0^2} + {\rm e}^{-\Delta_-^2/\Delta_0^2}}{8 \sqrt{\pi} \Delta_0}  \frac{ \Omega^{N - 3/2} \sqrt{1 - \Omega}}{\left( 1 + \alpha - \Omega \right)^2}  \:,\\
&&\Delta_{\pm} = - \bar{\Delta}(E_{\text{exc}}) \pm  \frac{\delta |2 \Omega - 1|}{2 \sqrt{\Omega (1- \Omega)}}\:, \nonumber
\end{eqnarray}
see details in SOM. The model adequately describes the Raman shift of the e-SFRS line, as shown in Fig.~\ref{fig3}(b). The analysis demonstrates that for $\alpha \ll 1$ the effective dispersion of the $g$ factors, leading to the inhomogeneous Raman line width, is much smaller than that of the experimental line, which is mainly determined by the spectral width of the monochromator slits.

In ideal bulk semiconductors the electron states from the $\Gamma$- and X-valleys do not mix with each other, however $\Gamma$-X mixing does take place in low-dimensional heterostructures due to reflection of the electron wave from the interfaces. For (001)-oriented GaAs/AlAs superlattices with type-II band alignment it has been demonstrated that strain and quantum confinement lift the level degeneracy at the X-valley of the AlAs layer~\cite{Fu,Voliotis}. Depending on the GaAs layer thickness the ground state can be either at the X$_z$ or X$_{xy}$ valley. In superlattices considerable $\Gamma$-X$_z$ mixing is provided by the uncertainty of the electron $k$ vector, $k_{z}$, perpendicular to the interface~\cite{Meynadier88,Li94}. The $\Gamma$-X$_{xy}$ mixing can be induced only by violation of the translational symmetry in the $xy$-plane, which is absent in superlattices with flat interfaces. However, this violation is possible in QDs due to their boundaries perpendicular to the (001)-direction. For the (In,Al)As/AlAs QDs this mechanism is responsible for the strong variation of the exciton recombination rate that is affected by the annealing treatment during growth~\cite{Shamirzaevnew}, which in turn changes the $\Gamma$-X$_{xy}$ mixing. SFRS is therefore able to characterize the $\Gamma$-X$_{xy}$ mixing quantitatively, opening up a novel way for systematic studies of inter-valley coupling and spin-flip scattering processes in semiconductors.

To conclude, we have shown that the spin level structure of the indirect-in-momentum-space exciton in type-I (In,Al)As/AlAs QDs can be assessed by resonant spin-flip Raman scattering due to its mixing with the optically allowed direct exciton. This tool can be also applied to other indirect systems with, e.g., $\Gamma$-L-valley mixed excitons or type-II band alignment. Moreover, our study implies that it is worthwhile to further attempt tailoring of the band structure of such mixed direct-indirect systems, as in that way one can obtain long-lived excitons with appealing spin properties that do not only have long relaxation times but can also be manipulated by optical or electrical methods. These QD structures are promising for quantum information technologies. Also, the SFRS itself is a coherent manipulation, as it can be used to initialize and orient the spins of the electron, heavy-hole, and in particular the exciton.

This work was supported by the Deutsche Forschungsgemeinschaft via SPP 1285, Mercator Research Center Ruhr (MERCUR) of Stiftung Mercator, Russian Foundation for Basic Research (Grants No. 13-02-00073 and  14-02-00033), Russian Ministry of Education and Science (Contract No. 11.G34.31.0067), and Russian Federation President Grant NSh-1085.2014.2.


\begin{thebibliography}{}
\bibitem{QDbook} \textit{Lasers Based on Quantum Dot Structures: Physics and Devices}, E.~U.~Rafailov \textit{et al.} (Wiley-VCH, Weinheim, 2011).
\bibitem{Cortez} S.~Cortez \textit{et al.}, Phys. Rev. Lett. \textbf{89}, 207401 (2002).
\bibitem{Kammerer} C.~Kammerer \textit{et al.}, Phys. Rev. B \textbf{66}, 041306(R) (2002).
\bibitem{Kroutvar} M.~Kroutvar \textit{et al.}, Nature \textbf{432}, 81 (2004).
\bibitem{Senes} M.~S\'{e}n\`{e}s \textit{et al.}, Phys. Rev. B \textbf{71}, 115334 (2005).
\bibitem{Yu} H.~Yu \textit{et al.}, Appl. Phys. Lett. \textbf{69}, 4087 (1996).
\bibitem{Awschalom} D.~D.~Awschalom \textit{et al.}, Nature Phys. \textbf{3}, 153 (2007).
\bibitem{Vuckovic} J.~Vuckovic \textit{et al.}, Appl. Phys. Lett. \textbf{82}, 2374 (2003).
\bibitem{Kress} A.~Kress \textit{et al.}, Phys. Rev. B \textbf{71}, 241304(R) (2005).
\bibitem{Gershoni} E.~Poem \textit{et al.}, Nature Phys. \textbf{6}, 993 (2010).
\bibitem{Shamirzaevb} T.~S.~Shamirzaev \textit{et al.}, Phys. Rev. B \textbf{78}, 085323 (2008).
\bibitem{Shamirzaevnew} T.~S.~Shamirzaev \textit{et al.}, Phys. Rev. B \textbf{84}, 155318 (2011).
\bibitem{Koudinov} A.~V.~Koudinov \textit{et al.}, Phys. Rev. B \textbf{79}, 241310(R) (2009).
\bibitem{Debus} J.~Debus \textit{et al.}, Phys. Rev. B \textbf{87}, 205316 (2013).
\bibitem{Sirenko} A.~A.~Sirenko \textit{et al.}, Phys. Rev. B \textbf{58}, 2077 (1998).
\bibitem{Puls} J.~Puls \textit{et al.}, Phys. Rev. B \textbf{60}, R16303 (1999).
\bibitem{Shamirzaev} T.~S.~Shamirzaev \textit{et al.}, Appl. Phys. Lett. \textbf{92}, 213101 (2008).
\bibitem{Luo} J.-W.~Luo \textit{et al.}, Phys. Rev. B \textbf{78}, 035306 (2008).
\bibitem{commentrange} The crossing energy of the $\Gamma$- and X-levels is spread over the $1.6$$-$$1.7$~eV energy range due to QD parameter variations.
\bibitem{Young} C.~F.~Young \textit{et al.}, Phys. Rev. B \textbf{55}, 16245 (1997).
\bibitem{Vdovin} E.~E.~Vdovin \textit{et al.}, Phys. Rev. B \textbf{71}, 195320 (2005).
\bibitem{Yugovab} I.~A.~Yugova \textit{et al.}, Phys. Rev. B \textbf{75}, 245302 (2007).
\bibitem{Sapega} V.~F.~Sapega \textit{et al.}, Phys. Rev. B \textbf{45}, 4320 (1992).
\bibitem{Faradaygeom} We also performed Raman scattering experiments in the Faraday geometry and did not find any SFRS signal, as expected. Electron and heavy-hole spin-flips are forbidden in the electric-dipole approximation for high potential symmetries~\cite{Debus}. On the other hand, the exciton spin-flip is allowed in crossed circular polarizations, but cannot be resolved due to the rather small Zeeman splitting.
\bibitem{HPL} I.~Ya.~Karlik \textit{et al.}, Sov. Phys. Semicond. \textbf{21}, 630 (1987).
\bibitem{comment1} The absence of a further e-line in the SFRS spectrum of Fig.~\ref{fig2}(a) suggests that the value of $g_{_\Gamma}$ in the studied dot structures is small and hereafter we set it equal to zero for simplicity. One could account in the theoretical model for a small but finite $g_{_\Gamma}$ value. It would not significantly change the calculation results presented here, however the equations become rather bulky.
\bibitem{Nazarov} A.~V.~Khaetskii \textit{et al.}, Phys. Rev. B \textbf{64}, 125316 (2001).
\bibitem{remark} For resonant excitation the exciton recombination via non-radiative channels is assumed to be slow: $\tau_{\text{nr}} \gg \tau_{_\Gamma}$.
\bibitem{Fu} Y.~Fu \textit{et al.}, Phys. Rev. B \textbf{47}, 13498 (1993).
\bibitem{Voliotis} V.~Voliotis \textit{et al.}, Phys. Rev. B \textbf{49}, 2576 (1994).
\bibitem{Meynadier88} M.-H.~Meynadier \textit{et al.}, Phys. Rev. Lett. \textbf{60}, 1338 (1988).
\bibitem{Li94} G.~H.~Li \textit{et al.}, Phys. Rev. B \textbf{50}, 18420 (1994).
\end{thebibliography}
\end{document}